\documentclass[amsmath,amssymb,twocolumn,noeprint]{revtex4-1}
\usepackage[utf8]{inputenc}
\usepackage{graphicx}
\usepackage[percent]{overpic}
\usepackage{color}
\usepackage{amsmath}
\usepackage{amsfonts} 
\usepackage{amssymb}
\newcommand\comment[1]{} 
\begin{document}
\title{Out-of-plane dielectric susceptibility of graphene in twistronic and Bernal bilayers}

\author{Sergey Slizovskiy} 
\altaffiliation{on leave from NRC ``Kurchatov Institute'' PNPI, Russia}
\author{Aitor Garcia-Ruiz}
\author{Alexey I. Berdyugin}
\author{Na Xin}
\affiliation{National Graphene Institute, University of Manchester, Booth St.E., M13 9PL, Manchester, UK} 
\affiliation{Dept.\ of Physics \& Astronomy, University of Manchester, Manchester M13 9PL, UK}
\author{Takashi Taniguchi}
\affiliation{National Institute for Materials Science, 1-1 Namiki, Tsukuba, 305-0044, Japan}
\author{Kenji Watanabe}
\affiliation{National Institute for Materials Science, 1-1 Namiki, Tsukuba, 305-0044, Japan}
\author{Andre K. Geim}
\affiliation{National Graphene Institute,, University of Manchester, Booth St.E., M13 9PL, Manchester, UK} 
\affiliation{Department of Physics \& Astronomy, University of Manchester, Manchester M13 9PL, UK}
\author{ Neil\ D.\ Drummond}
\affiliation{Department of Physics, Lancaster University, Lancaster LA1 4YB, UK} 
\author{Vladimir I. Fal’ko}
\email{vladimir.falko@manchester.ac.uk}
\affiliation{National Graphene Institute, Booth St.E., M13 9PL, Manchester, UK} 
\affiliation{Department of Physics \& Astronomy, University of Manchester, Manchester M13 9PL, UK}
\affiliation{Henry Royce Institute for Advanced Materials, Manchester, M13 9PL, UK}

\date{\today}
\begin{abstract}
We describe how the out-of-plane dielectric polarizability of monolayer graphene influences the electrostatics of bilayer graphene – both Bernal (BLG) and twisted (tBLG). We compare the polarizability value computed using density functional theory with the output from previously published experimental data on the electrostatically controlled interlayer asymmetry potential in BLG and data on the on-layer density distribution in tBLG.  We show that monolayers in tBLG are described well by polarizability $\alpha_{exp} = 10.8 \,\mathrm{\AA}^3$ and effective out-of-plane dielectric susceptibility $\epsilon_z = 2.5$, including their on-layer electron density distribution at zero magnetic field and the inter-layer Landau level pinning at quantizing magnetic fields. 
\end{abstract}

\maketitle

Bilayer graphene \cite{NP2006,McCann_t006-2,McCann_t006-1} is a two-dimensional (2D) material with electronic properties tuneable over a broad range. The manifestations of the qualitative change of electronic characteristics of both Bernal (BLG) and twisted (tBLG) bilayer graphene, produced by  electrostatic gating \cite{McCann_t006-1} and inter-layer misalignment \cite{CastroNeto2007,Bistritzer_2011}, were observed in numerous experimental studies of the electronic transport in graphene-based field-effect transistor (FET) devices. These versatile electronic properties make FETs based on BLG and tBLG an attractive hardware platform for applications tailored \cite{NP2007,Ensslin_2019,Efetov_2020} for various quantum technologies. While, over the recent years, the fundamental electronic properties of bilayer graphene have been intensively studied, a mundane but practical characteristic of this material related to the out-of-plane dielectric susceptibility of graphene layers largely escaped attention of those investigations of BLG and tBLG in FETs, despite several already recorded indications \cite{Sanchez-Yamagishi_2012,Fallahazad_2012,arXiv2019,Ensslin2020,Focusing2020} of its relevance for the quantitative modelling of the operation of such devices. 

The out-of-plane dielectric susceptibility of a single graphene layer stems from the polarisability of its carbon orbitals, that is, from the mixing of $\pi$ and $\sigma$ bands by an electric field oriented perpendicular to the 2D crystal. Here, we compute the effective dielectric susceptibility, $\epsilon_z$, of graphene monolayer using {\it ab initio} density functional theory (DFT) and implement the estimated DFT value of $\epsilon_z$ in the self-consistent description of (a) tBLG electrostatics in tBLG with twist angles outside the magic angle \cite{Bistritzer_2011,Efetov_2020} range and (b) conditions for the inter-layer Landau level pinning in FET with a $30^\circ$-twisted bilayer, with the results of modelling showing favourable quantitative agreement with the available and new experimental data. Then we take into account the out-of-plane dielectric susceptibility of a single graphene layer in the analysis of BLG inter-layer asymmetry gap, in particular, its dependence on the vertical displacement field, $\Delta (D)$, comparing the results with the measured exciton spectroscopy in gapped BLG \cite{Ju_2017}. 

 For the theoretical modelling of the out-of-plane dielectric susceptibility, we employ the CASTEP plane-wave-basis DFT code \cite{castep} with ultra-soft pseudopotentials. We use a 53×53×1 $k$-point grid, a large plane-wave cut-off of 566 eV, and a variety of interlayer distances $c$ along the $z$-axis to compute the total energy, ${\cal E}$, of graphene in a saw-tooth potential, $-Dz/\epsilon_0$, centered on the carbon sites of the graphene layer ($D$ being the displacement field and $-c/2<z<c/2$). Then we determine \footnote{Note that, at larger external fields, the energy abruptly becomes non-quadratic in $D$ due to electronic density appearing in the artificial triangular well of the saw-tooth potential, which sets the limits for the applicability of the DFT method we used. Also, we find that $\alpha$ is sensitive to the plane-wave cut-off energy at small external fields, which limits from below the range of $D$ values we used in the analysis. We verified that the same polarizability results were obtained by directly evaluating the change in the dipole moment within the simulation cell when the external field is applied. Note that here we differs from some earlier studies of, e.g., bilayers \cite{OldDFT, DFT2013}, where the dielectric screening contribution has not been separated from the contribution resulting from charge redistribution across the layers.} the polarizability $\alpha$ in each cell of length $c$ using the relation ${\cal E}  ={\cal E}_0-\alpha D^2/(2\epsilon_0 )$, with ${\cal E}_0$ being the vacuum energy.  As the artificial periodicity, introduced in the DFT code, leads to a systematic error in the polarizability, $\delta \alpha (c) \propto c^{-1}$, we fit the obtained DFT data with $\alpha(c)=\alpha_{\infty}+a/c+b/c^2$ and find $\alpha_{DFT} \equiv \alpha(c \rightarrow \infty) = 11.0 \,\mathrm{\AA}^3$  per unit cell of graphene with the Perdew--Burke--Ernzerhof (PBE) functional and  $\alpha_{DFT}=10.8 \,\mathrm{\AA}^3$ with the local density approximation (LDA).  These values are close to the DFT-PBE polarizability reported in Ref.\ \onlinecite{OldDFT}, $\alpha = 0.867 \times 4 \pi \,\mathrm{\AA}^3 = 10.9 \,\mathrm{\AA}^3$, and when  recalculated into an effective 'electronic thickness' $\alpha / \mathcal{A}$, where  $\mathcal{A}=5.2 \, \rm \AA^2$ is graphene's unit cell area, we get $2.1 \, \mathrm{\AA}$, comparable to the earlier-quoted `electronic thickness' of graphene \cite{Microscopic16,Ensslin2020}. Furthermore, for the analysis of bilayers' electrostatics in this paper, the effective out-of-plane dielectric susceptibility is $\epsilon_z=[ 1-\alpha_{DFT}/\mathcal{A} d ] ^{-1}$, where $d$ is the distance between the carbon planes in the bilayer ($d=3.35\, \rm \AA$ for BLG and $d=3.44\, \rm \AA$ for tBLG, as in turbostratic graphite \cite{Turbostratic90}). This gives $\epsilon_z = 2.6$ for BLG, and $\epsilon_z = 2.5$ for tBLG.

\begin{figure}
\begin{center}
\includegraphics[width=0.99 \columnwidth]{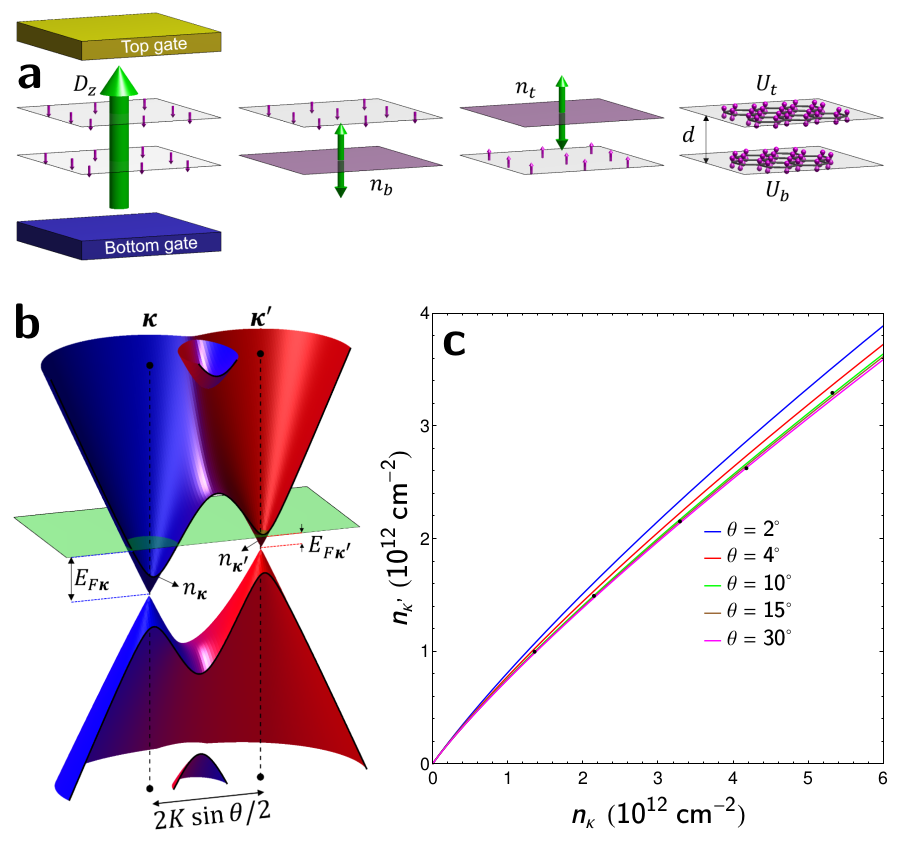}
\caption{(a) Sketches illustrating how dielectric polarizability of each monolayer enters in the electrostatics analysis of bilayers in Eq.\ (\ref{eq:U2mU1}).  (b) Characteristic electron dispersion in tBLG (here, $\theta = 3^\circ$; $u = 100 \,\rm meV$). Electron state amplitude on the top/bottom layer is shown by red/blue. (c) Mini-valley carrier densities $n_{\kappa/\kappa'}$ in a single-gated tBLG calculated for various misalignment angles outside the magic angle range, in comparison with the densities corresponding to SdHO measured \cite{Fallahazad_2012} in a tBLG flake with an unknown twist angle (black dots). 
\label{fig:SingleGate}}
\end{center}
\end{figure}

 For the electrostatic analysis of bilayers built into FETs, we note that out-of-plane polarisation of carbon orbitals in each graphene monolayer is decoupled from the charges hosted by its own $\pi$-bands, because of mirror-symmetric charge and field distribution produced by the latter, see in Fig.\ \ref{fig:SingleGate}(a). As a result, the difference between the on-layer potential energies in the top and bottom layers of a bilayer, each with the electron density $n_{b/t}$, has the form,
\begin{equation}\label{eq:U2mU1}
u  \equiv  U_t-U_b    = e
\left[
\frac{D}{\epsilon_0\epsilon_z}-
e\frac{1+\epsilon_z^{-1}}{2 \epsilon_0}
\left(
\frac{n_b-n_t}{2}
\right)
\right]d.
\end{equation}
This expression is applicable to the description of both BLG and tBLG in a FET, improving on the earlier-published studies \cite{McCann_t006-1,Review2009,McCann_t013} where the out-of-plane dielectric susceptibility of graphene layers was missed out in the self-consistent band structure analysis. 

To describe a twisted bilayer with an interlayer twist angle $\theta$, we use the minimal tBLG Hamiltonian, \cite{CastroNeto2007,Bistritzer_2011}, 
\begin{align}\label{eq:H_tBLG}
\mathcal{H}(\boldsymbol{k}',\boldsymbol{k})=&
\left(
\begin{matrix}
[\mathcal{H}_t + \frac12 u]\delta_{\boldsymbol{k}',\boldsymbol{k}} &
\mathcal{T}_{\boldsymbol{k}',\boldsymbol{k}}\\
\mathcal{T}_{\boldsymbol{k},\boldsymbol{k}'}^\dagger&
[\mathcal{H}_b - \frac12 u]\delta_{\boldsymbol{k}',\boldsymbol{k}}
\end{matrix}
\right),\\
\mathcal{H}_{t/b}=& 
\hbar v \left(
\begin{matrix}
0 & \pi_{t/b,\xi}^{*}\\
\pi_{t/b,\xi} & 0
\end{matrix}
\right),\nonumber
\\
\pi_{t/b,\xi}=&\,\xi k_x+i(k_y\mp \xi K \sin \frac{\theta}{2} ),
\nonumber
\\
\mathcal{T}_{\boldsymbol{k}',\boldsymbol{k}}=&
\frac{\gamma_1}{3}
\sum_{j=0}^2
\left(
\begin{matrix}
1&e^{i\xi\frac{2\pi}{3}j}\\
e^{-i\xi\frac{2\pi}{3}j}&1
\end{matrix}
\right)
\delta_{\boldsymbol{k}',\boldsymbol{k}+\boldsymbol{g}_\xi^{(j)}}, \nonumber \\
\boldsymbol{g}_\xi^{(j)} = &\, \xi \left(-\sin\frac{2\pi j}{3},1-\cos\frac{2\pi j}{3}\right) 2K\sin\frac{\theta}{2} .
\nonumber
\end{align}
Here, $v = 10.2\cdot 10^6 \,\rm m/sec$ is Dirac velocity in monolayer graphene. Equation (\ref{eq:H_tBLG}) determines \cite{CastroNeto2007} characteristic low-energy bands, illustrated in Fig.\ \ref{fig:SingleGate}(b) for  $1 \gg \theta \ge \gamma_1/\hbar v K \equiv 2^{\circ}$ (away from the small magic angles $\le 1^\circ$). This spectrum features two Dirac minivalleys at $\boldsymbol{\kappa}$ and $\boldsymbol{\kappa}'$ ($|\boldsymbol{\kappa}-\boldsymbol{\kappa}'| = 2K \sin \frac{\theta}{2}$), each described by its own Fermi energy, 
\begin{align}
& E_{F\, \boldsymbol{\kappa}/\boldsymbol{\kappa}'} \approx (1-3 \lambda) \hbar v \sqrt{\pi |n_{\boldsymbol{\kappa}/\boldsymbol{\kappa}'}|} {\,\mathrm{sign}\,} n_{\boldsymbol{\kappa}/\boldsymbol{\kappa}'},  \nonumber
\\ 
& E_{F\, \boldsymbol{\kappa}} - E_{F\, \boldsymbol{\kappa}'}  \approx  (1-4 \lambda)u ,  
\label{tBLG1}
\end{align}
and carrier density $n_{\boldsymbol{\kappa}/\boldsymbol{\kappa}'}$, determined by the minivalley area encircled by the corresponding Fermi lines (as in Fig.\ \ref{fig:SingleGate}(b)) and measurable using Shubnikov - de Haas oscillations \cite{Fallahazad_2012,Sanchez-Yamagishi_2012} or Fabry-Perot interference pattern \cite{Ensslin2020}. 
The above expression was obtained using expansion up to the linear order in $\lambda = [\frac{\gamma_1}{\hbar v \,|\boldsymbol{\kappa}-\boldsymbol{\kappa}'|}]^2 \ll 1$. We also note that, due to the interlayer hybridisation of electronic wave functions, the on-layer charge densities in Eq.\ (\ref{eq:U2mU1}) differ from the minivalley carrier densities, 
\begin{equation}
n_{b/t} \approx n_{\boldsymbol{\kappa}/\boldsymbol{\kappa}'} \pm 2 \lambda \left(n_{\boldsymbol{\kappa}'}-n_{\boldsymbol{\kappa}} + 0.07 \frac{u}{\hbar v} |\boldsymbol{\kappa}-\boldsymbol{\kappa}'| \right),
\label{tBLG2}
\end{equation}
which makes the results of the self-consistent analysis of tBLG electrostatics slightly dependent on the twist angle, $\theta$. We illustrate this weak dependence in Fig. \ref{fig:SingleGate}(c) by plotting the relation between the values of $n_{\kappa}$ and $n_{\kappa'}$ in a single-side-gated tBLG computed using Eqs. (\ref{tBLG1}), (\ref{tBLG2}) and (\ref{eq:U2mU1}) with $\epsilon_z = 2.5$ and $d=3.44\, \rm \AA$. On the same plot, we also show the values of $n_{\boldsymbol{\kappa}}$ and $n_{\boldsymbol{\kappa}'}$ recalculated from the earlier measured SdHO \cite{Fallahazad_2012} in tBLG devices with an unknown twist angle (seemingly, $10^{\circ}-15^{\circ}$).  

\begin{figure*}
\begin{center}
\includegraphics[width=1.99 \columnwidth]{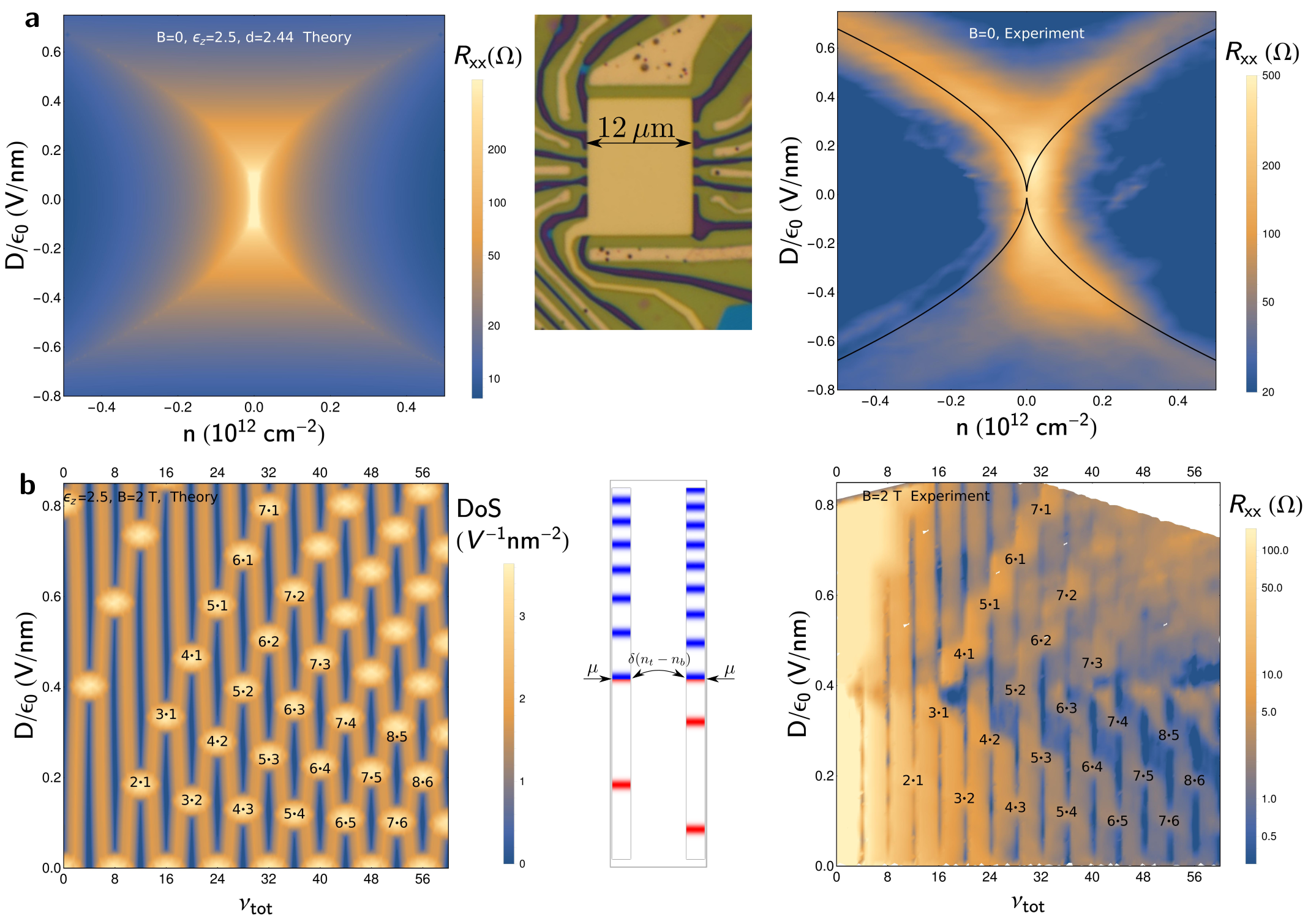}
\caption{(a) Resistance map for a double-gated tBLG with a $30^\circ$ twist angle, computed with $\epsilon_z = 2.5$ and $d=3.44\, \rm \AA$ (left) and measured (right) as a function of the total carrier density, $n$ and vertical displacement field, $D$, at $B=0$ and $T=2 \,$K. (b) Computed density of states of pinning LLs (left) and the measured resistance, $\rho_{xx}$ (right) in a $30^\circ$ tBLG at $B=2$ Tesla, plotted as a function of displacement field and filling factor. Bright regions correspond to the marked $N_t / N_b$ LL pinning conditions.
\label{fig:LLs2}}
\end{center}
\end{figure*}

We also compared the results of the self-consistent tBLG analysis with the measurements of electronic transport characteristics of a double-side gated multi-terminal FET based on a tBLG with $\theta = 30^\circ$. The latter choice provides us with the maximum misalignment, corresponding to $\lambda \rightarrow 0$ in Eqs.\ (\ref{tBLG1}) and (\ref{tBLG2}), hence $n_{b/t}=n_{\boldsymbol{\kappa}/\boldsymbol{\kappa}'}$. In the experimentally studied device, tBLG was encapsulated between hBN films on the top and bottom, thus providing both a precise electrostatic control of tBLG for $B=0$ measurements and its high-mobility, enabling to observe quantum Hall effect at a magnetic field as low as $B=2$ Tesla. 

For a quantitative comparison of the measured and modelled tBLG characteristics, we assumed elastic scattering of carriers from residual Coulomb impurities in the encapsulating environment with a dielectric constant $\epsilon \approx 5$, with an areal density $n_c$, screened jointly by the carriers in the top and bottom layers. The screening determines \cite{Cheianov_2006} the Fourier form-factor of the scatterers, 
\begin{equation}\label{u} 
\phi_q = \frac{e^2 / 2\epsilon_0 \epsilon}{q + r_s (k_{Ft} + k_{Fb})}, \,\, r_s = \frac{e^2/\epsilon_0 \epsilon }{\pi \hbar v} ,  k_{Fi}=\sqrt{\pi|n_{i}|}, 
\nonumber
\end{equation}
and the corresponding momentum relaxation rate of Dirac electrons \cite{Cheianov_2006},
$$\tau_{t/b}^{-1} = \frac{\pi \gamma_{t/b}}{2\hbar} \langle |\phi_{2 k_{Ft/b} \sin(\varphi/2)}|^2 \sin^2\varphi \rangle_\varphi n_c.$$
Then, in Fig.\ \ref{fig:LLs2}(a), we compare the computed and measured tBLG resistivity. As in monolayer graphene \cite{Cheianov_2006}, density of states, $\gamma_{t/b}$, cancels out from each $\rho_{t/b} = 2\tau_{t/b}^{-1}/ (e^2v^2\gamma_{t/b})$, making the overall result, $\rho_{xx} = \rho_t \rho_b / [\rho_t + \rho_b]$, dependent on the carrier density only through the wave-number transfer, $2 k_{Ft/b} \sin(\varphi/2)$, and screening. This produces ridge-like resistance maxima at $k_{Ft}=0$ or $k_{Fb}=0$, that is, when 
\begin{equation}\label{D-n}
    \pm D/\epsilon_0  = \frac{\hbar v \epsilon_z \sqrt{\pi |n|}{\rm\, sign\,} n}{e d} + \frac{e\, n (1+ \epsilon_z)}{4 \epsilon_0}.
\end{equation}
Lines corresponding to the above relation are laid over the experimentally measured resistivity map for a direct comparison.    

We find an even more compelling coincidence between the theory and experiment when studying the Landau level pinning between two electronically independent by electrostatically coupled graphene monolayers in a $30^\circ$ tBLG. In a magnetic field, graphene spectrum splits into Landau levels (LLs) with energies $E_N =  v  \sqrt{2 \hbar|  e\, B\, N|} \, {\rm sign} N$. In a twisted bilayer, infinite degeneracy of LLs gives a leeway to the interlayer charge transfer which screens out displacement field and pins partially filled top/bottom layer LLs, $N_t$ and $N_b$, to each other and to their common chemical potential, $\mu$, so that $v \sqrt{2 \hbar e|B|} (\sqrt{N_t} - \sqrt{N_b}) = u$, as in Fig.\ \ref{fig:LLs2}(b). 
This LL pining effect also persists for slightly broadened LLs. Taking into account a small Gaussian LL broadening, $\Gamma$, we write,
\begin{equation}
n_{t/b} = \frac{4 e B}{2 \pi \hbar}\sum_{N=-\Lambda}^\Lambda {\rm erf}\left(\frac{\mu \pm \frac12 u - E_N}{\Gamma}  \right), \ \ \Lambda \gg 1,  
\end{equation}
solve self-consistently Eq.\ (\ref {eq:U2mU1}), and compute the total density of states (DoS) in the bilayer. The computed DoS for $B=2$ Tesla and $\Gamma \sim 0.5 \,\rm meV$) is mapped in Fig.\ \ref{fig:LLs2}(b) versus displacement field and tBLG filling factor, $\nu_{tot}=hn/eB$. Here, the `bright' high-DoS spots indicate the interlayer LL pinning conditions, whereas the `dark' low-DoS streaks mark conditions for incompressible states in a tBLG.  We compare this map with $\rho_{xx}(D,\nu_{tot})$ measured in the quantum Hall effect regimes (similar to the ones observed earlier \cite{Sanchez-Yamagishi_2012, Fallahazad_2012} in other tBLG devices), where the high resistivity manifests mutual pinning of partially filled LLs, whereas the minima correspond to the incompressible states. To mention, the computed pattern broadly varies upon changing $\epsilon_z$, whereas the value of $\epsilon_z=2.5$ gives an excellent match between the computed and measured maps in Fig.\ \ref{fig:LLs2}(b).

\begin{figure}
\begin{center}
\includegraphics[width=1\columnwidth]{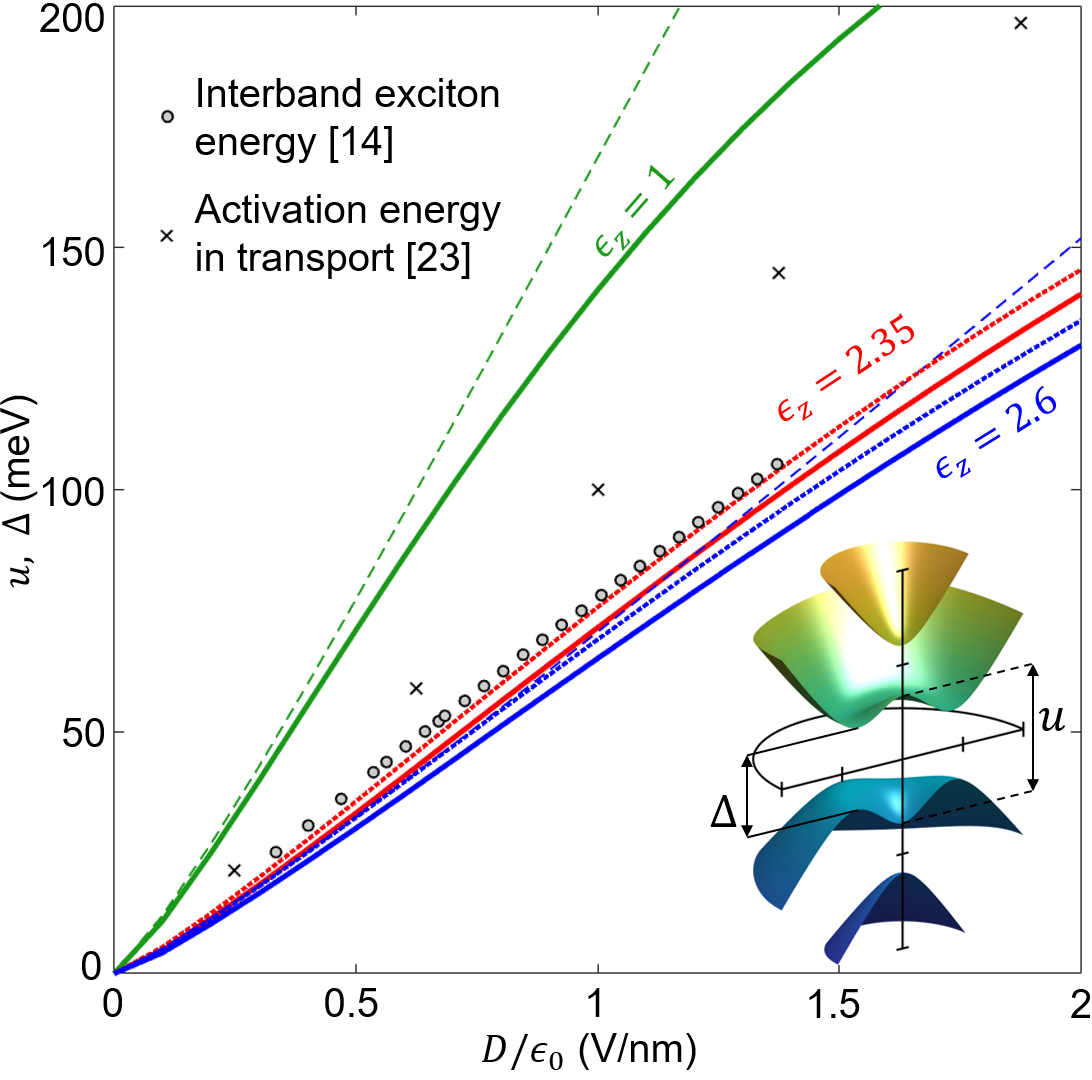}
\caption{Inter-layer asymmetry potential (dashed lines) and band gap (solid lines) in an undoped BLG, self-consistently computed with various values of $\epsilon_z =$ 1 (green), 2.6 (blue) and 2.35 (red) and compared to the experimentally measured gate-tunable optical gap \cite{Ju_2017} (circles) and transport gap \cite{Zhang_2009} (crosses). Here, we use \cite{Kuzmenko_2009, NegativeGamma3} $v = 10.2\cdot 10^6 \,\rm m/sec$, $\gamma_1= 0.38$\,eV, $v_{3}=1.23\cdot10^5$\,m/s,  $v_4=4.54\cdot10^{4}$\,m/s, $\delta=22$\,meV, and $d=3.35\, \rm \AA$. Additionally, dotted lines show the values of the gap computed with $\gamma_1= 0.35$\,eV  and the same all other parameters.  Sketch illustrates 4 BLG bands (1 and 2 below / 3 and 4 above the gap) highlighting a small difference between $u$ and $\Delta$. \label{fig:Delta}}
\end{center}
\end{figure}

Finally, we analyse the electrostatically controlled asymmetry gap \cite{McCann_t006-2} in Bernal bilayer graphene, taking into account out-of-plane polarizability of its constituent monolayers. In this case, we use Eq. (\ref{eq:U2mU1}) with $\epsilon_z = [ 1-\alpha/\mathcal{A} d ] ^{-1} \approx 2.6$, recalculated from polarizability $\alpha_{DFT}$ using $d=3.35 \, \AA$, and the BLG Hamiltonian \cite{McCann_t006-2}, 
\begin{align}
\cal{H}=
\left(
\begin{matrix}
\frac{u}{2} &v\hbar\pi^*_\xi&-v_4\hbar\pi_\xi^*&-v_3\hbar\pi_\xi\\
v\hbar\pi_\xi&\delta+\frac{u}{2}&\gamma_1&-v_4\hbar\pi_\xi^*\\
-v_4\hbar\pi_\xi&\gamma_1&\delta-\frac{u}{2}&v\hbar\pi^*_\xi\\
-v_3\hbar\pi_\xi^*&-v_4\hbar\pi_\xi&v\hbar\pi_\xi^*&-\frac{u}{2}
\end{matrix}
\right), 
\label{H_BLG}
\end{align}
which determines the dispersion and the sublattice (A/B) amplitudes, $\Psi_{\lambda, t/b,\boldsymbol{k}}^{\beta}$, in four ($\beta = 1-4$) spin- and valley-degenerate bands, $E_{\boldsymbol{k}}^\beta$. Here, $\pi_\xi\equiv \xi k_x+ik_y$, $\boldsymbol{k}=(k_x,k_y)$ is the electron wave vector in the valleys $\boldsymbol{K}_{\xi}=\xi(4\pi/3a,0)$, $\xi = \pm$. The on-layer electron densities in an undoped gapped BLG (with Fermi level in the gap between bands $\beta = 1,2$ and $\beta = 3,4$) are
\begin{align}
n_{t/b}=
\int
\frac{d^2\boldsymbol{k}}{\pi^2}
\sum_{\beta=1,2}
\left[
\sum_{\lambda = A/B}
|\Psi_{\lambda, t/b,\boldsymbol{k}}^{\beta}|^2 - \frac{1}{4} 
\right],  
\label{n_BLG}
\end{align}

The on-layer potential energy difference, $u$, and a band gap, $\Delta$ in the BLG spectrum (see inset in Fig. \ref{fig:Delta}), computed using self-consistent analysis of Eqs.\ (\ref{H_BLG}), (\ref{n_BLG}) and (\ref{eq:U2mU1}) with $\epsilon_z=1$ (as in Refs.\ \onlinecite{McCann_t006-1,Review2009,McCann_t013}) and with $\epsilon_z=2.6$, are plotted in Fig.\ \ref{fig:Delta} versus displacement field, $D$. On the same plot, we show the values of activation energy in lateral transport \cite{BLGGapZhang08} and the IR 'optical gap' - interlayer exciton energy \cite{Ju_2017}, measured in various BLG devices. The difference between those two data sets is due to that the single-electron 'transport' gap is enhanced by the self-energy correction \cite{Cheianov_2012} due to the electron-electron repulsion, as compared to the `electrostatic' value, $u$, whereas that enhancement is mostly cancelled out by the binding energy of the exciton \cite{Cheianov_2012,Ju_2017}, an optically active electron-hole bound state. As one can see in Fig. \ref{fig:Delta}, $u$ and $\Delta$ computed without taking into account monolayer's polarizability ($\epsilon_z=1$) largely overestimate their values. At the same time, the values of $u$ and $\Delta$ obtained using $\epsilon_z=2.6$ appear to be less than the exciton energy measured in optics, for interlayer coupling across the whole range $0.35$\,eV\,$<\gamma_1 <$\,0.38\,eV covered in the previous literature \cite{Kuzmenko_2009, BLGGapMak2009,Wirth21,AbInitMacDonald,JungMacDonald, NegativeGamma3,CR2021}.  This discrepancy may be related to that the interaction terms in the electron self-energy are only partially cancelled by the exciton binding energy \cite{Cheianov_2012}. It may also signal that the out-of-plane monolayer polarizability, $\alpha$, is reduced by $\sim 10$\% when it is part of BLG, as the values of $\Delta$ computed with $\epsilon_z=2.35$ and $\gamma_1 = 0.35$\,eV agree very well with the measured optical gap values.         

In summary, the reported analysis of the out-of-plane dielectric susceptibility of monolayer graphene shows that the latter plays an important role in determining the electrostatics of both Bernal and twisted bilayer graphene. We found that the DFT-computed polarizability of the monolayer, $\alpha = 10.8 \,\mathrm{\AA}^3$, accounts very well for all details of the electrostatics of twisted bilayers, including the on-layer electron density distribution at zero magnetic field and the inter-layer Landau level pinning at quantizing magnetic fields. For practical applications in modeling of FET devices based on twisted bilayers, the polarizability of monolayer graphene can be converted to its effective dielectric susceptibility, $\epsilon_z = 2.5$, which should be used for the self-consistent electrostatic analysis of tBLG using Eq.\ (\ref{eq:U2mU1}) of this manuscript. 

This work was supported by EPSRC grants EP/S019367/1, EP/S030719/1, EP/N010345/1, ERC Synergy Grant Hetero2D, Lloyd Register Foundation Nanotechnology grant, and the European Graphene Flagship Core 3 Project. Computing resources were provided by Lancaster HEC cluster and Manchester SCF.


\begin{thebibliography}{33}%
\makeatletter
\providecommand \@ifxundefined [1]{%
 \@ifx{#1\undefined}
}%
\providecommand \@ifnum [1]{%
 \ifnum #1\expandafter \@firstoftwo
 \else \expandafter \@secondoftwo
 \fi
}%
\providecommand \@ifx [1]{%
 \ifx #1\expandafter \@firstoftwo
 \else \expandafter \@secondoftwo
 \fi
}%
\providecommand \natexlab [1]{#1}%
\providecommand \enquote  [1]{``#1''}%
\providecommand \bibnamefont  [1]{#1}%
\providecommand \bibfnamefont [1]{#1}%
\providecommand \citenamefont [1]{#1}%
\providecommand \href@noop [0]{\@secondoftwo}%
\providecommand \href [0]{\begingroup \@sanitize@url \@href}%
\providecommand \@href[1]{\@@startlink{#1}\@@href}%
\providecommand \@@href[1]{\endgroup#1\@@endlink}%
\providecommand \@sanitize@url [0]{\catcode `\\12\catcode `\$12\catcode
  `\&12\catcode `\#12\catcode `\^12\catcode `\_12\catcode `\%12\relax}%
\providecommand \@@startlink[1]{}%
\providecommand \@@endlink[0]{}%
\providecommand \url  [0]{\begingroup\@sanitize@url \@url }%
\providecommand \@url [1]{\endgroup\@href {#1}{\urlprefix }}%
\providecommand \urlprefix  [0]{URL }%
\providecommand \Eprint [0]{\href }%
\providecommand \doibase [0]{http://dx.doi.org/}%
\providecommand \selectlanguage [0]{\@gobble}%
\providecommand \bibinfo  [0]{\@secondoftwo}%
\providecommand \bibfield  [0]{\@secondoftwo}%
\providecommand \translation [1]{[#1]}%
\providecommand \BibitemOpen [0]{}%
\providecommand \bibitemStop [0]{}%
\providecommand \bibitemNoStop [0]{.\EOS\space}%
\providecommand \EOS [0]{\spacefactor3000\relax}%
\providecommand \BibitemShut  [1]{\csname bibitem#1\endcsname}%
\let\auto@bib@innerbib\@empty
\bibitem [{\citenamefont {{Novoselov}}\ \emph {et~al.}(2006)\citenamefont
  {{Novoselov}}, \citenamefont {{McCann}}, \citenamefont {{Morozov}},
  \citenamefont {{Fal'ko}}, \citenamefont {{Katsnelson}}, \citenamefont
  {{Zeitler}}, \citenamefont {{Jiang}}, \citenamefont {{Schedin}},\ and\
  \citenamefont {{Geim}}}]{NP2006}%
  \BibitemOpen
  \bibfield  {author} {\bibinfo {author} {\bibfnamefont {K.~S.}\ \bibnamefont
  {{Novoselov}}}, \bibinfo {author} {\bibfnamefont {E.}~\bibnamefont
  {{McCann}}}, \bibinfo {author} {\bibfnamefont {S.~V.}\ \bibnamefont
  {{Morozov}}}, \bibinfo {author} {\bibfnamefont {V.~I.}\ \bibnamefont
  {{Fal'ko}}}, \bibinfo {author} {\bibfnamefont {M.~I.}\ \bibnamefont
  {{Katsnelson}}}, \bibinfo {author} {\bibfnamefont {U.}~\bibnamefont
  {{Zeitler}}}, \bibinfo {author} {\bibfnamefont {D.}~\bibnamefont {{Jiang}}},
  \bibinfo {author} {\bibfnamefont {F.}~\bibnamefont {{Schedin}}}, \ and\
  \bibinfo {author} {\bibfnamefont {A.~K.}\ \bibnamefont {{Geim}}},\ }\href
  {\doibase 10.1038/nphys245} {\bibfield  {journal} {\bibinfo  {journal}
  {Nature Physics}\ }\textbf {\bibinfo {volume} {2}},\ \bibinfo {pages} {177}
  (\bibinfo {year} {2006})}\BibitemShut {NoStop}%
\bibitem [{\citenamefont {McCann}\ and\ \citenamefont
  {Fal'ko}(2006)}]{McCann_t006-2}%
  \BibitemOpen
  \bibfield  {author} {\bibinfo {author} {\bibfnamefont {E.}~\bibnamefont
  {McCann}}\ and\ \bibinfo {author} {\bibfnamefont {V.~I.}\ \bibnamefont
  {Fal'ko}},\ }\href {\doibase 10.1103/PhysRevLett.96.086805} {\bibfield
  {journal} {\bibinfo  {journal} {Phys. Rev. Lett.}\ }\textbf {\bibinfo
  {volume} {96}},\ \bibinfo {pages} {086805} (\bibinfo {year}
  {2006})}\BibitemShut {NoStop}%
\bibitem [{\citenamefont {McCann}(2006)}]{McCann_t006-1}%
  \BibitemOpen
  \bibfield  {author} {\bibinfo {author} {\bibfnamefont {E.}~\bibnamefont
  {McCann}},\ }\href {\doibase 10.1103/PhysRevB.74.161403} {\bibfield
  {journal} {\bibinfo  {journal} {Phys. Rev. B}\ }\textbf {\bibinfo {volume}
  {74}},\ \bibinfo {pages} {161403} (\bibinfo {year} {2006})}\BibitemShut
  {NoStop}%
\bibitem [{\citenamefont {Lopes~dos Santos}\ \emph {et~al.}(2007)\citenamefont
  {Lopes~dos Santos}, \citenamefont {Peres},\ and\ \citenamefont
  {Castro~Neto}}]{CastroNeto2007}%
  \BibitemOpen
  \bibfield  {author} {\bibinfo {author} {\bibfnamefont {J.~M.~B.}\
  \bibnamefont {Lopes~dos Santos}}, \bibinfo {author} {\bibfnamefont
  {N.~M.~R.}\ \bibnamefont {Peres}}, \ and\ \bibinfo {author} {\bibfnamefont
  {A.~H.}\ \bibnamefont {Castro~Neto}},\ }\href {\doibase
  10.1103/PhysRevLett.99.256802} {\bibfield  {journal} {\bibinfo  {journal}
  {Phys. Rev. Lett.}\ }\textbf {\bibinfo {volume} {99}},\ \bibinfo {pages}
  {256802} (\bibinfo {year} {2007})}\BibitemShut {NoStop}%
\bibitem [{\citenamefont {Bistritzer}\ and\ \citenamefont
  {MacDonald}(2011)}]{Bistritzer_2011}%
  \BibitemOpen
  \bibfield  {author} {\bibinfo {author} {\bibfnamefont {R.}~\bibnamefont
  {Bistritzer}}\ and\ \bibinfo {author} {\bibfnamefont {A.~H.}\ \bibnamefont
  {MacDonald}},\ }\href {\doibase 10.1073/pnas.1108174108} {\bibfield
  {journal} {\bibinfo  {journal} {Proceedings of the National Academy of
  Sciences}\ }\textbf {\bibinfo {volume} {108}},\ \bibinfo {pages} {12233}
  (\bibinfo {year} {2011})}\BibitemShut {NoStop}%
\bibitem [{\citenamefont {Falko}(2007)}]{NP2007}%
  \BibitemOpen
  \bibfield  {author} {\bibinfo {author} {\bibfnamefont {V.}~\bibnamefont
  {Falko}},\ }\href {\doibase 10.1038/nphys556} {\bibfield  {journal} {\bibinfo
   {journal} {Nature Physics}\ }\textbf {\bibinfo {volume} {3}},\ \bibinfo
  {pages} {151} (\bibinfo {year} {2007})}\BibitemShut {NoStop}%
\bibitem [{\citenamefont {Kurzmann}\ \emph {et~al.}(2019)\citenamefont
  {Kurzmann}, \citenamefont {Overweg}, \citenamefont {Eich}, \citenamefont
  {Pally}, \citenamefont {Rickhaus}, \citenamefont {Pisoni}, \citenamefont
  {Lee}, \citenamefont {Watanabe}, \citenamefont {Taniguchi}, \citenamefont
  {Ihn},\ and\ \citenamefont {Ensslin}}]{Ensslin_2019}%
  \BibitemOpen
  \bibfield  {author} {\bibinfo {author} {\bibfnamefont {A.}~\bibnamefont
  {Kurzmann}}, \bibinfo {author} {\bibfnamefont {H.}~\bibnamefont {Overweg}},
  \bibinfo {author} {\bibfnamefont {M.}~\bibnamefont {Eich}}, \bibinfo {author}
  {\bibfnamefont {A.}~\bibnamefont {Pally}}, \bibinfo {author} {\bibfnamefont
  {P.}~\bibnamefont {Rickhaus}}, \bibinfo {author} {\bibfnamefont
  {R.}~\bibnamefont {Pisoni}}, \bibinfo {author} {\bibfnamefont
  {Y.}~\bibnamefont {Lee}}, \bibinfo {author} {\bibfnamefont {K.}~\bibnamefont
  {Watanabe}}, \bibinfo {author} {\bibfnamefont {T.}~\bibnamefont {Taniguchi}},
  \bibinfo {author} {\bibfnamefont {T.}~\bibnamefont {Ihn}}, \ and\ \bibinfo
  {author} {\bibfnamefont {K.}~\bibnamefont {Ensslin}},\ }\href {\doibase
  10.1021/acs.nanolett.9b01617} {\bibfield  {journal} {\bibinfo  {journal}
  {Nano Letters}\ }\textbf {\bibinfo {volume} {19}},\ \bibinfo {pages} {5216}
  (\bibinfo {year} {2019})}\BibitemShut {NoStop}%
\bibitem [{\citenamefont {{Seifert}}\ \emph {et~al.}(2020)\citenamefont
  {{Seifert}}, \citenamefont {{Lu}}, \citenamefont {{Stepanov}}, \citenamefont
  {{Dur{\'a}n Retamal}}, \citenamefont {{Moore}}, \citenamefont {{Fong}},
  \citenamefont {{Principi}},\ and\ \citenamefont {{Efetov}}}]{Efetov_2020}%
  \BibitemOpen
  \bibfield  {author} {\bibinfo {author} {\bibfnamefont {P.}~\bibnamefont
  {{Seifert}}}, \bibinfo {author} {\bibfnamefont {X.}~\bibnamefont {{Lu}}},
  \bibinfo {author} {\bibfnamefont {P.}~\bibnamefont {{Stepanov}}}, \bibinfo
  {author} {\bibfnamefont {J.~R.}\ \bibnamefont {{Dur{\'a}n Retamal}}},
  \bibinfo {author} {\bibfnamefont {J.~N.}\ \bibnamefont {{Moore}}}, \bibinfo
  {author} {\bibfnamefont {K.-C.}\ \bibnamefont {{Fong}}}, \bibinfo {author}
  {\bibfnamefont {A.}~\bibnamefont {{Principi}}}, \ and\ \bibinfo {author}
  {\bibfnamefont {D.~K.}\ \bibnamefont {{Efetov}}},\ }\href {\doibase
  10.1021/acs.nanolett.0c00373} {\bibfield  {journal} {\bibinfo  {journal}
  {Nano Letters}\ }\textbf {\bibinfo {volume} {20}},\ \bibinfo {pages} {3459}
  (\bibinfo {year} {2020})}\BibitemShut {NoStop}%
\bibitem [{\citenamefont {Sanchez-Yamagishi}\ \emph {et~al.}(2012)\citenamefont
  {Sanchez-Yamagishi}, \citenamefont {Taychatanapat}, \citenamefont {Watanabe},
  \citenamefont {Taniguchi}, \citenamefont {Yacoby},\ and\ \citenamefont
  {Jarillo-Herrero}}]{Sanchez-Yamagishi_2012}%
  \BibitemOpen
  \bibfield  {author} {\bibinfo {author} {\bibfnamefont {J.~D.}\ \bibnamefont
  {Sanchez-Yamagishi}}, \bibinfo {author} {\bibfnamefont {T.}~\bibnamefont
  {Taychatanapat}}, \bibinfo {author} {\bibfnamefont {K.}~\bibnamefont
  {Watanabe}}, \bibinfo {author} {\bibfnamefont {T.}~\bibnamefont {Taniguchi}},
  \bibinfo {author} {\bibfnamefont {A.}~\bibnamefont {Yacoby}}, \ and\ \bibinfo
  {author} {\bibfnamefont {P.}~\bibnamefont {Jarillo-Herrero}},\ }\href
  {\doibase 10.1103/PhysRevLett.108.076601} {\bibfield  {journal} {\bibinfo
  {journal} {Phys. Rev. Lett.}\ }\textbf {\bibinfo {volume} {108}},\ \bibinfo
  {pages} {076601} (\bibinfo {year} {2012})}\BibitemShut {NoStop}%
\bibitem [{\citenamefont {Fallahazad}\ \emph {et~al.}(2012)\citenamefont
  {Fallahazad}, \citenamefont {Hao}, \citenamefont {Lee}, \citenamefont {Kim},
  \citenamefont {Ruoff},\ and\ \citenamefont {Tutuc}}]{Fallahazad_2012}%
  \BibitemOpen
  \bibfield  {author} {\bibinfo {author} {\bibfnamefont {B.}~\bibnamefont
  {Fallahazad}}, \bibinfo {author} {\bibfnamefont {Y.}~\bibnamefont {Hao}},
  \bibinfo {author} {\bibfnamefont {K.}~\bibnamefont {Lee}}, \bibinfo {author}
  {\bibfnamefont {S.}~\bibnamefont {Kim}}, \bibinfo {author} {\bibfnamefont
  {R.~S.}\ \bibnamefont {Ruoff}}, \ and\ \bibinfo {author} {\bibfnamefont
  {E.}~\bibnamefont {Tutuc}},\ }\href {\doibase 10.1103/PhysRevB.85.201408}
  {\bibfield  {journal} {\bibinfo  {journal} {Phys. Rev. B}\ }\textbf {\bibinfo
  {volume} {85}},\ \bibinfo {pages} {201408} (\bibinfo {year}
  {2012})}\BibitemShut {NoStop}%
\bibitem [{\citenamefont {Slizovskiy}\ \emph {et~al.}(2019)\citenamefont
  {Slizovskiy}, \citenamefont {Garcia-Ruiz}, \citenamefont {Drummond},\ and\
  \citenamefont {Fal'ko}}]{arXiv2019}%
  \BibitemOpen
  \bibfield  {author} {\bibinfo {author} {\bibfnamefont {S.}~\bibnamefont
  {Slizovskiy}}, \bibinfo {author} {\bibfnamefont {A.}~\bibnamefont
  {Garcia-Ruiz}}, \bibinfo {author} {\bibfnamefont {N.}~\bibnamefont
  {Drummond}}, \ and\ \bibinfo {author} {\bibfnamefont {V.~I.}\ \bibnamefont
  {Fal'ko}},\ }\href@noop {} {\enquote {\bibinfo {title} {Dielectric
  susceptibility of graphene describing its out-of-plane polarizability},}\ }
  (\bibinfo {year} {2019})\BibitemShut {NoStop}%
\bibitem [{\citenamefont {Rickhaus}\ \emph {et~al.}(2020)\citenamefont
  {Rickhaus}, \citenamefont {Liu}, \citenamefont {Kurpas}, \citenamefont
  {Kurzmann}, \citenamefont {Lee}, \citenamefont {Overweg}, \citenamefont
  {Eich}, \citenamefont {Pisoni}, \citenamefont {Taniguchi}, \citenamefont
  {Watanabe}, \citenamefont {Richter}, \citenamefont {Ensslin},\ and\
  \citenamefont {Ihn}}]{Ensslin2020}%
  \BibitemOpen
  \bibfield  {author} {\bibinfo {author} {\bibfnamefont {P.}~\bibnamefont
  {Rickhaus}}, \bibinfo {author} {\bibfnamefont {M.-H.}\ \bibnamefont {Liu}},
  \bibinfo {author} {\bibfnamefont {M.}~\bibnamefont {Kurpas}}, \bibinfo
  {author} {\bibfnamefont {A.}~\bibnamefont {Kurzmann}}, \bibinfo {author}
  {\bibfnamefont {Y.}~\bibnamefont {Lee}}, \bibinfo {author} {\bibfnamefont
  {H.}~\bibnamefont {Overweg}}, \bibinfo {author} {\bibfnamefont
  {M.}~\bibnamefont {Eich}}, \bibinfo {author} {\bibfnamefont {R.}~\bibnamefont
  {Pisoni}}, \bibinfo {author} {\bibfnamefont {T.}~\bibnamefont {Taniguchi}},
  \bibinfo {author} {\bibfnamefont {K.}~\bibnamefont {Watanabe}}, \bibinfo
  {author} {\bibfnamefont {K.}~\bibnamefont {Richter}}, \bibinfo {author}
  {\bibfnamefont {K.}~\bibnamefont {Ensslin}}, \ and\ \bibinfo {author}
  {\bibfnamefont {T.}~\bibnamefont {Ihn}},\ }\href {\doibase
  10.1126/sciadv.aay8409} {\bibfield  {journal} {\bibinfo  {journal} {Science
  Advances}\ }\textbf {\bibinfo {volume} {6}},\ \bibinfo {pages} {eaay8409}
  (\bibinfo {year} {2020})}\BibitemShut {NoStop}%
\bibitem [{\citenamefont {Berdyugin}\ \emph {et~al.}(2020)\citenamefont
  {Berdyugin}, \citenamefont {Tsim}, \citenamefont {Kumaravadivel},
  \citenamefont {Xu}, \citenamefont {Ceferino}, \citenamefont {Knothe},
  \citenamefont {Kumar}, \citenamefont {Taniguchi}, \citenamefont {Watanabe},
  \citenamefont {Geim}, \citenamefont {Grigorieva},\ and\ \citenamefont
  {Fal'ko}}]{Focusing2020}%
  \BibitemOpen
  \bibfield  {author} {\bibinfo {author} {\bibfnamefont {A.~I.}\ \bibnamefont
  {Berdyugin}}, \bibinfo {author} {\bibfnamefont {B.}~\bibnamefont {Tsim}},
  \bibinfo {author} {\bibfnamefont {P.}~\bibnamefont {Kumaravadivel}}, \bibinfo
  {author} {\bibfnamefont {S.~G.}\ \bibnamefont {Xu}}, \bibinfo {author}
  {\bibfnamefont {A.}~\bibnamefont {Ceferino}}, \bibinfo {author}
  {\bibfnamefont {A.}~\bibnamefont {Knothe}}, \bibinfo {author} {\bibfnamefont
  {R.~K.}\ \bibnamefont {Kumar}}, \bibinfo {author} {\bibfnamefont
  {T.}~\bibnamefont {Taniguchi}}, \bibinfo {author} {\bibfnamefont
  {K.}~\bibnamefont {Watanabe}}, \bibinfo {author} {\bibfnamefont {A.~K.}\
  \bibnamefont {Geim}}, \bibinfo {author} {\bibfnamefont {I.~V.}\ \bibnamefont
  {Grigorieva}}, \ and\ \bibinfo {author} {\bibfnamefont {V.~I.}\ \bibnamefont
  {Fal'ko}},\ }\href {\doibase 10.1126/sciadv.aay7838} {\bibfield  {journal}
  {\bibinfo  {journal} {Science Advances}\ }\textbf {\bibinfo {volume} {6}},\
  \bibinfo {pages} {eaay7838} (\bibinfo {year} {2020})}\BibitemShut {NoStop}%
\bibitem [{\citenamefont {Ju}\ \emph {et~al.}(2017)\citenamefont {Ju},
  \citenamefont {Wang}, \citenamefont {Cao}, \citenamefont {Taniguchi},
  \citenamefont {Watanabe}, \citenamefont {Louie}, \citenamefont {Rana},
  \citenamefont {Park}, \citenamefont {Hone}, \citenamefont {Wang},\ and\
  \citenamefont {McEuen}}]{Ju_2017}%
  \BibitemOpen
  \bibfield  {author} {\bibinfo {author} {\bibfnamefont {L.}~\bibnamefont
  {Ju}}, \bibinfo {author} {\bibfnamefont {L.}~\bibnamefont {Wang}}, \bibinfo
  {author} {\bibfnamefont {T.}~\bibnamefont {Cao}}, \bibinfo {author}
  {\bibfnamefont {T.}~\bibnamefont {Taniguchi}}, \bibinfo {author}
  {\bibfnamefont {K.}~\bibnamefont {Watanabe}}, \bibinfo {author}
  {\bibfnamefont {S.~G.}\ \bibnamefont {Louie}}, \bibinfo {author}
  {\bibfnamefont {F.}~\bibnamefont {Rana}}, \bibinfo {author} {\bibfnamefont
  {J.}~\bibnamefont {Park}}, \bibinfo {author} {\bibfnamefont {J.}~\bibnamefont
  {Hone}}, \bibinfo {author} {\bibfnamefont {F.}~\bibnamefont {Wang}}, \ and\
  \bibinfo {author} {\bibfnamefont {P.~L.}\ \bibnamefont {McEuen}},\ }\href
  {\doibase 10.1126/science.aam9175} {\bibfield  {journal} {\bibinfo  {journal}
  {Science}\ }\textbf {\bibinfo {volume} {358}},\ \bibinfo {pages} {907}
  (\bibinfo {year} {2017})}\BibitemShut {NoStop}%
\bibitem [{\citenamefont {{Clark}}\ \emph {et~al.}(2005)\citenamefont
  {{Clark}}, \citenamefont {{Segall}}, \citenamefont {{Pickard}}, \citenamefont
  {{Hasnip}}, \citenamefont {{Probert}}, \citenamefont {{Refson}},\ and\
  \citenamefont {{Payne}}}]{castep}%
  \BibitemOpen
  \bibfield  {author} {\bibinfo {author} {\bibfnamefont {S.~J.}\ \bibnamefont
  {{Clark}}}, \bibinfo {author} {\bibfnamefont {M.~D.}\ \bibnamefont
  {{Segall}}}, \bibinfo {author} {\bibfnamefont {C.~J.}\ \bibnamefont
  {{Pickard}}}, \bibinfo {author} {\bibfnamefont {P.~J.}\ \bibnamefont
  {{Hasnip}}}, \bibinfo {author} {\bibfnamefont {M.~I.~J.}\ \bibnamefont
  {{Probert}}}, \bibinfo {author} {\bibfnamefont {K.}~\bibnamefont {{Refson}}},
  \ and\ \bibinfo {author} {\bibfnamefont {M.~C.}\ \bibnamefont {{Payne}}},\
  }\href {\doibase 10.1524/zkri.220.5.567.65075} {\bibfield  {journal}
  {\bibinfo  {journal} {Zeitschrift fur Kristallographie}\ }\textbf {\bibinfo
  {volume} {220}},\ \bibinfo {pages} {567} (\bibinfo {year}
  {2005})}\BibitemShut {NoStop}%
\bibitem [{Note1()}]{Note1}%
  \BibitemOpen
  \bibinfo {note} {Note that, at larger external fields, the energy abruptly
  becomes non-quadratic in $D$ due to electronic density appearing in the
  artificial triangular well of the saw-tooth potential, which sets the limits
  for the applicability of the DFT method we used. Also, we find that $\alpha $
  is sensitive to the plane-wave cut-off energy at small external fields, which
  limits from below the range of $D$ values we used in the analysis. We
  verified that the same polarizability results were obtained by directly
  evaluating the change in the dipole moment within the simulation cell when
  the external field is applied. Note that here we differs from some earlier
  studies of, e.g., bilayers \cite {OldDFT, DFT2013}, where the dielectric
  screening contribution has not been separated from the contribution resulting
  from charge redistribution across the layers.}\BibitemShut {Stop}%
\bibitem [{\citenamefont {Yu}\ \emph {et~al.}(2008)\citenamefont {Yu},
  \citenamefont {Stewart},\ and\ \citenamefont {Tiwari}}]{OldDFT}%
  \BibitemOpen
  \bibfield  {author} {\bibinfo {author} {\bibfnamefont {E.~K.}\ \bibnamefont
  {Yu}}, \bibinfo {author} {\bibfnamefont {D.~A.}\ \bibnamefont {Stewart}}, \
  and\ \bibinfo {author} {\bibfnamefont {S.}~\bibnamefont {Tiwari}},\ }\href
  {\doibase 10.1103/PhysRevB.77.195406} {\bibfield  {journal} {\bibinfo
  {journal} {Phys. Rev. B}\ }\textbf {\bibinfo {volume} {77}},\ \bibinfo
  {pages} {195406} (\bibinfo {year} {2008})}\BibitemShut {NoStop}%
\bibitem [{\citenamefont {Fang}\ \emph {et~al.}(2016)\citenamefont {Fang},
  \citenamefont {Vandenberghe},\ and\ \citenamefont
  {Fischetti}}]{Microscopic16}%
  \BibitemOpen
  \bibfield  {author} {\bibinfo {author} {\bibfnamefont {J.}~\bibnamefont
  {Fang}}, \bibinfo {author} {\bibfnamefont {W.~G.}\ \bibnamefont
  {Vandenberghe}}, \ and\ \bibinfo {author} {\bibfnamefont {M.~V.}\
  \bibnamefont {Fischetti}},\ }\href {\doibase 10.1103/PhysRevB.94.045318}
  {\bibfield  {journal} {\bibinfo  {journal} {Phys. Rev. B}\ }\textbf {\bibinfo
  {volume} {94}},\ \bibinfo {pages} {045318} (\bibinfo {year}
  {2016})}\BibitemShut {NoStop}%
\bibitem [{\citenamefont {Bayot}\ \emph {et~al.}(1990)\citenamefont {Bayot},
  \citenamefont {Piraux}, \citenamefont {Michenaud}, \citenamefont {Issi},
  \citenamefont {Lelaurain},\ and\ \citenamefont {Moore}}]{Turbostratic90}%
  \BibitemOpen
  \bibfield  {author} {\bibinfo {author} {\bibfnamefont {V.}~\bibnamefont
  {Bayot}}, \bibinfo {author} {\bibfnamefont {L.}~\bibnamefont {Piraux}},
  \bibinfo {author} {\bibfnamefont {J.-P.}\ \bibnamefont {Michenaud}}, \bibinfo
  {author} {\bibfnamefont {J.-P.}\ \bibnamefont {Issi}}, \bibinfo {author}
  {\bibfnamefont {M.}~\bibnamefont {Lelaurain}}, \ and\ \bibinfo {author}
  {\bibfnamefont {A.}~\bibnamefont {Moore}},\ }\href {\doibase
  10.1103/PhysRevB.41.11770} {\bibfield  {journal} {\bibinfo  {journal} {Phys.
  Rev. B}\ }\textbf {\bibinfo {volume} {41}},\ \bibinfo {pages} {11770}
  (\bibinfo {year} {1990})}\BibitemShut {NoStop}%
\bibitem [{\citenamefont {Castro~Neto}\ \emph {et~al.}(2009)\citenamefont
  {Castro~Neto}, \citenamefont {Guinea}, \citenamefont {Peres}, \citenamefont
  {Novoselov},\ and\ \citenamefont {Geim}}]{Review2009}%
  \BibitemOpen
  \bibfield  {author} {\bibinfo {author} {\bibfnamefont {A.~H.}\ \bibnamefont
  {Castro~Neto}}, \bibinfo {author} {\bibfnamefont {F.}~\bibnamefont {Guinea}},
  \bibinfo {author} {\bibfnamefont {N.~M.~R.}\ \bibnamefont {Peres}}, \bibinfo
  {author} {\bibfnamefont {K.~S.}\ \bibnamefont {Novoselov}}, \ and\ \bibinfo
  {author} {\bibfnamefont {A.~K.}\ \bibnamefont {Geim}},\ }\href {\doibase
  10.1103/RevModPhys.81.109} {\bibfield  {journal} {\bibinfo  {journal} {Rev.
  Mod. Phys.}\ }\textbf {\bibinfo {volume} {81}},\ \bibinfo {pages} {109}
  (\bibinfo {year} {2009})}\BibitemShut {NoStop}%
\bibitem [{\citenamefont {McCann}\ and\ \citenamefont
  {Koshino}(2013)}]{McCann_t013}%
  \BibitemOpen
  \bibfield  {author} {\bibinfo {author} {\bibfnamefont {E.}~\bibnamefont
  {McCann}}\ and\ \bibinfo {author} {\bibfnamefont {M.}~\bibnamefont
  {Koshino}},\ }\href {\doibase 10.1088/0034-4885/76/5/056503} {\bibfield
  {journal} {\bibinfo  {journal} {Reports on Progress in Physics}\ }\textbf
  {\bibinfo {volume} {76}},\ \bibinfo {pages} {56503} (\bibinfo {year}
  {2013})}\BibitemShut {NoStop}%
\bibitem [{\citenamefont {Cheianov}\ and\ \citenamefont
  {Fal'ko}(2006)}]{Cheianov_2006}%
  \BibitemOpen
  \bibfield  {author} {\bibinfo {author} {\bibfnamefont {V.~V.}\ \bibnamefont
  {Cheianov}}\ and\ \bibinfo {author} {\bibfnamefont {V.~I.}\ \bibnamefont
  {Fal'ko}},\ }\href {\doibase 10.1103/PhysRevLett.97.226801} {\bibfield
  {journal} {\bibinfo  {journal} {Phys. Rev. Lett.}\ }\textbf {\bibinfo
  {volume} {97}},\ \bibinfo {pages} {226801} (\bibinfo {year}
  {2006})}\BibitemShut {NoStop}%
\bibitem [{\citenamefont {{Zhang}}\ \emph {et~al.}(2009)\citenamefont
  {{Zhang}}, \citenamefont {{Tang}}, \citenamefont {{Girit}}, \citenamefont
  {{Hao}}, \citenamefont {{Martin}}, \citenamefont {{Zettl}}, \citenamefont
  {{Crommie}}, \citenamefont {{Shen}},\ and\ \citenamefont
  {{Wang}}}]{Zhang_2009}%
  \BibitemOpen
  \bibfield  {author} {\bibinfo {author} {\bibfnamefont {Y.}~\bibnamefont
  {{Zhang}}}, \bibinfo {author} {\bibfnamefont {T.-T.}\ \bibnamefont {{Tang}}},
  \bibinfo {author} {\bibfnamefont {C.}~\bibnamefont {{Girit}}}, \bibinfo
  {author} {\bibfnamefont {Z.}~\bibnamefont {{Hao}}}, \bibinfo {author}
  {\bibfnamefont {M.~C.}\ \bibnamefont {{Martin}}}, \bibinfo {author}
  {\bibfnamefont {A.}~\bibnamefont {{Zettl}}}, \bibinfo {author} {\bibfnamefont
  {M.~F.}\ \bibnamefont {{Crommie}}}, \bibinfo {author} {\bibfnamefont {Y.~R.}\
  \bibnamefont {{Shen}}}, \ and\ \bibinfo {author} {\bibfnamefont
  {F.}~\bibnamefont {{Wang}}},\ }\href {\doibase 10.1038/nature08105}
  {\bibfield  {journal} {\bibinfo  {journal} {\nat}\ }\textbf {\bibinfo
  {volume} {459}},\ \bibinfo {pages} {820} (\bibinfo {year}
  {2009})}\BibitemShut {NoStop}%
\bibitem [{\citenamefont {Kuzmenko}\ \emph {et~al.}(2009)\citenamefont
  {Kuzmenko}, \citenamefont {Crassee},\ and\ \citenamefont {van~der
  Marel}}]{Kuzmenko_2009}%
  \BibitemOpen
  \bibfield  {author} {\bibinfo {author} {\bibfnamefont {A.~B.}\ \bibnamefont
  {Kuzmenko}}, \bibinfo {author} {\bibfnamefont {I.}~\bibnamefont {Crassee}}, \
  and\ \bibinfo {author} {\bibfnamefont {D.}~\bibnamefont {van~der Marel}},\
  }\href {\doibase DOI 10.1103/PhysRevB.80.165406} {\bibfield  {journal}
  {\bibinfo  {journal} {Physical Review B}\ }\textbf {\bibinfo {volume} {80,
  165406}},\ \bibinfo {pages} {165406} (\bibinfo {year} {2009})}\BibitemShut
  {NoStop}%
\bibitem [{\citenamefont {Joucken}\ \emph {et~al.}(2020)\citenamefont
  {Joucken}, \citenamefont {Ge}, \citenamefont {Quezada-L\'opez}, \citenamefont
  {Davenport}, \citenamefont {Watanabe}, \citenamefont {Taniguchi},\ and\
  \citenamefont {Velasco}}]{NegativeGamma3}%
  \BibitemOpen
  \bibfield  {author} {\bibinfo {author} {\bibfnamefont {F.}~\bibnamefont
  {Joucken}}, \bibinfo {author} {\bibfnamefont {Z.}~\bibnamefont {Ge}},
  \bibinfo {author} {\bibfnamefont {E.~A.}\ \bibnamefont {Quezada-L\'opez}},
  \bibinfo {author} {\bibfnamefont {J.~L.}\ \bibnamefont {Davenport}}, \bibinfo
  {author} {\bibfnamefont {K.}~\bibnamefont {Watanabe}}, \bibinfo {author}
  {\bibfnamefont {T.}~\bibnamefont {Taniguchi}}, \ and\ \bibinfo {author}
  {\bibfnamefont {J.}~\bibnamefont {Velasco}},\ }\href {\doibase
  10.1103/PhysRevB.101.161103} {\bibfield  {journal} {\bibinfo  {journal}
  {Phys. Rev. B}\ }\textbf {\bibinfo {volume} {101}},\ \bibinfo {pages}
  {161103} (\bibinfo {year} {2020})}\BibitemShut {NoStop}%
\bibitem [{\citenamefont {Zhang}\ \emph {et~al.}(2008)\citenamefont {Zhang},
  \citenamefont {Li}, \citenamefont {Basov}, \citenamefont {Fogler},
  \citenamefont {Hao},\ and\ \citenamefont {Martin}}]{BLGGapZhang08}%
  \BibitemOpen
  \bibfield  {author} {\bibinfo {author} {\bibfnamefont {L.~M.}\ \bibnamefont
  {Zhang}}, \bibinfo {author} {\bibfnamefont {Z.~Q.}\ \bibnamefont {Li}},
  \bibinfo {author} {\bibfnamefont {D.~N.}\ \bibnamefont {Basov}}, \bibinfo
  {author} {\bibfnamefont {M.~M.}\ \bibnamefont {Fogler}}, \bibinfo {author}
  {\bibfnamefont {Z.}~\bibnamefont {Hao}}, \ and\ \bibinfo {author}
  {\bibfnamefont {M.~C.}\ \bibnamefont {Martin}},\ }\href {\doibase
  10.1103/PhysRevB.78.235408} {\bibfield  {journal} {\bibinfo  {journal} {Phys.
  Rev. B}\ }\textbf {\bibinfo {volume} {78}},\ \bibinfo {pages} {235408}
  (\bibinfo {year} {2008})}\BibitemShut {NoStop}%
\bibitem [{\citenamefont {Cheianov}\ \emph {et~al.}(2012)\citenamefont
  {Cheianov}, \citenamefont {Aleiner},\ and\ \citenamefont
  {Fal'ko}}]{Cheianov_2012}%
  \BibitemOpen
  \bibfield  {author} {\bibinfo {author} {\bibfnamefont {V.~V.}\ \bibnamefont
  {Cheianov}}, \bibinfo {author} {\bibfnamefont {I.~L.}\ \bibnamefont
  {Aleiner}}, \ and\ \bibinfo {author} {\bibfnamefont {V.~I.}\ \bibnamefont
  {Fal'ko}},\ }\href {\doibase 10.1103/PhysRevLett.109.106801} {\bibfield
  {journal} {\bibinfo  {journal} {Phys. Rev. Lett.}\ }\textbf {\bibinfo
  {volume} {109}},\ \bibinfo {pages} {106801} (\bibinfo {year}
  {2012})}\BibitemShut {NoStop}%
\bibitem [{\citenamefont {Mak}\ \emph {et~al.}(2009)\citenamefont {Mak},
  \citenamefont {Lui}, \citenamefont {Shan},\ and\ \citenamefont
  {Heinz}}]{BLGGapMak2009}%
  \BibitemOpen
  \bibfield  {author} {\bibinfo {author} {\bibfnamefont {K.~F.}\ \bibnamefont
  {Mak}}, \bibinfo {author} {\bibfnamefont {C.~H.}\ \bibnamefont {Lui}},
  \bibinfo {author} {\bibfnamefont {J.}~\bibnamefont {Shan}}, \ and\ \bibinfo
  {author} {\bibfnamefont {T.~F.}\ \bibnamefont {Heinz}},\ }\href {\doibase
  10.1103/PhysRevLett.102.256405} {\bibfield  {journal} {\bibinfo  {journal}
  {Phys. Rev. Lett.}\ }\textbf {\bibinfo {volume} {102}},\ \bibinfo {pages}
  {256405} (\bibinfo {year} {2009})}\BibitemShut {NoStop}%
\bibitem [{\citenamefont {Wirth}\ \emph {et~al.}(2021)\citenamefont {Wirth},
  \citenamefont {Linnenbank}, \citenamefont {Steinle}, \citenamefont
  {Banszerus}, \citenamefont {Icking}, \citenamefont {Stampfer}, \citenamefont
  {Giessen},\ and\ \citenamefont {Taubner}}]{Wirth21}%
  \BibitemOpen
  \bibfield  {author} {\bibinfo {author} {\bibfnamefont {K.~G.}\ \bibnamefont
  {Wirth}}, \bibinfo {author} {\bibfnamefont {H.}~\bibnamefont {Linnenbank}},
  \bibinfo {author} {\bibfnamefont {T.}~\bibnamefont {Steinle}}, \bibinfo
  {author} {\bibfnamefont {L.}~\bibnamefont {Banszerus}}, \bibinfo {author}
  {\bibfnamefont {E.}~\bibnamefont {Icking}}, \bibinfo {author} {\bibfnamefont
  {C.}~\bibnamefont {Stampfer}}, \bibinfo {author} {\bibfnamefont
  {H.}~\bibnamefont {Giessen}}, \ and\ \bibinfo {author} {\bibfnamefont
  {T.}~\bibnamefont {Taubner}},\ }\href {\doibase 10.1021/acsphotonics.0c01442}
  {\bibfield  {journal} {\bibinfo  {journal} {ACS Photonics}\ }\textbf
  {\bibinfo {volume} {8}},\ \bibinfo {pages} {418} (\bibinfo {year}
  {2021})}\BibitemShut {NoStop}%
\bibitem [{\citenamefont {Min}\ \emph {et~al.}(2007)\citenamefont {Min},
  \citenamefont {Sahu}, \citenamefont {Banerjee},\ and\ \citenamefont
  {MacDonald}}]{AbInitMacDonald}%
  \BibitemOpen
  \bibfield  {author} {\bibinfo {author} {\bibfnamefont {H.}~\bibnamefont
  {Min}}, \bibinfo {author} {\bibfnamefont {B.}~\bibnamefont {Sahu}}, \bibinfo
  {author} {\bibfnamefont {S.~K.}\ \bibnamefont {Banerjee}}, \ and\ \bibinfo
  {author} {\bibfnamefont {A.~H.}\ \bibnamefont {MacDonald}},\ }\href {\doibase
  10.1103/PhysRevB.75.155115} {\bibfield  {journal} {\bibinfo  {journal} {Phys.
  Rev. B}\ }\textbf {\bibinfo {volume} {75}},\ \bibinfo {pages} {155115}
  (\bibinfo {year} {2007})}\BibitemShut {NoStop}%
\bibitem [{\citenamefont {Jung}\ and\ \citenamefont
  {MacDonald}(2014)}]{JungMacDonald}%
  \BibitemOpen
  \bibfield  {author} {\bibinfo {author} {\bibfnamefont {J.}~\bibnamefont
  {Jung}}\ and\ \bibinfo {author} {\bibfnamefont {A.~H.}\ \bibnamefont
  {MacDonald}},\ }\href {\doibase 10.1103/PhysRevB.89.035405} {\bibfield
  {journal} {\bibinfo  {journal} {Phys. Rev. B}\ }\textbf {\bibinfo {volume}
  {89}},\ \bibinfo {pages} {035405} (\bibinfo {year} {2014})}\BibitemShut
  {NoStop}%
\bibitem [{\citenamefont {Candussio}\ \emph {et~al.}(2021)\citenamefont
  {Candussio}, \citenamefont {Durnev}, \citenamefont {Slizovskiy},
  \citenamefont {J\"otten}, \citenamefont {Keil}, \citenamefont {Bel'kov},
  \citenamefont {Yin}, \citenamefont {Yang}, \citenamefont {Son}, \citenamefont
  {Mishchenko}, \citenamefont {Fal'ko},\ and\ \citenamefont
  {Ganichev}}]{CR2021}%
  \BibitemOpen
  \bibfield  {author} {\bibinfo {author} {\bibfnamefont {S.}~\bibnamefont
  {Candussio}}, \bibinfo {author} {\bibfnamefont {M.~V.}\ \bibnamefont
  {Durnev}}, \bibinfo {author} {\bibfnamefont {S.}~\bibnamefont {Slizovskiy}},
  \bibinfo {author} {\bibfnamefont {T.}~\bibnamefont {J\"otten}}, \bibinfo
  {author} {\bibfnamefont {J.}~\bibnamefont {Keil}}, \bibinfo {author}
  {\bibfnamefont {V.~V.}\ \bibnamefont {Bel'kov}}, \bibinfo {author}
  {\bibfnamefont {J.}~\bibnamefont {Yin}}, \bibinfo {author} {\bibfnamefont
  {Y.}~\bibnamefont {Yang}}, \bibinfo {author} {\bibfnamefont {S.-K.}\
  \bibnamefont {Son}}, \bibinfo {author} {\bibfnamefont {A.}~\bibnamefont
  {Mishchenko}}, \bibinfo {author} {\bibfnamefont {V.}~\bibnamefont {Fal'ko}},
  \ and\ \bibinfo {author} {\bibfnamefont {S.~D.}\ \bibnamefont {Ganichev}},\
  }\href {\doibase 10.1103/PhysRevB.103.125408} {\bibfield  {journal} {\bibinfo
   {journal} {Phys. Rev. B}\ }\textbf {\bibinfo {volume} {103}},\ \bibinfo
  {pages} {125408} (\bibinfo {year} {2021})}\BibitemShut {NoStop}%
\bibitem [{\citenamefont {Santos}\ and\ \citenamefont
  {Kaxiras}(2013)}]{DFT2013}%
  \BibitemOpen
  \bibfield  {author} {\bibinfo {author} {\bibfnamefont {E.~J.~G.}\
  \bibnamefont {Santos}}\ and\ \bibinfo {author} {\bibfnamefont
  {E.}~\bibnamefont {Kaxiras}},\ }\href {\doibase 10.1021/nl303611v} {\bibfield
   {journal} {\bibinfo  {journal} {Nano Letters}\ }\textbf {\bibinfo {volume}
  {13}},\ \bibinfo {pages} {898} (\bibinfo {year} {2013})}\BibitemShut
  {NoStop}%
\end{thebibliography}
\end{document}